\documentclass[a4paper,11pt]{article}
\pdfoutput=1
\usepackage{jinstpub} 

\title{Silicon photomultipliers in Very High Energy gamma-ray astrophysics}

\author[a,1]{D. Guberman,\note{Corresponding author.}}
\author[a,b]{R. Paoletti}
\affiliation[a]{Istituto Nazionale di Fisica Nucleare, Sezione di Pisa, I-56126 Pisa, Italy}
\affiliation[b]{Universita di Siena, I-53100 Siena, Italy}

\emailAdd{daniel.guberman@pi.infn.it}

\abstract{In the last ten years silicon photomultipliers (SiPMs) have gained terrain in experiments and applications in which photomultiplier tubes have been the dominant photosensors during decades. Imaging Atmospheric Cherenkov Telescopes (IACTs) for very high energy (VHE, E$>$50~GeV) gamma-ray astronomy are experiencing the same process. Until now FACT was the only IACT using SiPMs. In the Cherenkov Telescope Array (CTA), the next-generation VHE gamma-ray observatory, at least 70 telescopes equipped with SiPMs are planned to be built. The first prototypes have already been constructed and are now being commissioned. Here we discuss some of the advantages and drawbacks of using SiPMs in VHE gamma-ray astronomy and provide a brief overview of different developments related to the use of SiPMs in IACTs.}

\keywords{Photon detectors for UV, visible and IR photons (solid-state); Gamma telescopes; Detectors for UV, visible and IR photons}

\proceeding{5$^{\text{th}}$ INFIERI \\
  2019\\
  Wuhan, China}

\begin{document}
\maketitle
\flushbottom

\section{Introduction} \label{sec:intro}
Gamma rays at Very High Energy (VHE, $E>50$~GeV) have been observed from a wide variety of sources (e.g. pulsars, supernova remnants, active galactic nuclei) and provide essential information to study cosmic-ray acceleration, dark matter or Lorentz invariance violation (see~\citep{DeNaurois2015} for a recent review). Their energy and incoming direction can be retrieved by imaging the air showers they induce in the atmosphere using ground-based instruments such as water Cherenkov detectors located at high altitude (like HAWC~\citep{hawc_2013}) or using Imaging Atmospheric Cherenkov Telescopes (IACTs). IACTs, like VERITAS \citep{VERITAS2008}, H.E.S.S\citep{HESS2006} or MAGIC\citep{upgrade1}, are optical telescopes that image the showers by collecting the Cherenkov light flashes produced during the development of the shower. These telescopes feature $\sim$1-30~m diameter mirrors and cameras of a few hundreds to a few thousands pixels. The detection of the Cherenkov flashes is challenging because of their short duration (a few nanoseconds) and their low light intensity (down to a few photoelectrons per pixel).

IACT cameras have been equipped with photomultiplier tubes (PMTs) since the very beginning. With recent developments in silicon photomultipliers (SiPMs) there is a general trend for which experiments from different fields (high-energy physics, medical imaging) that traditionally employed PMTs switched, are switching or are considering to switch to SIPMs when possible. VHE gamma-ray astronomy is not an exception. Compared to PMTs, SiPMs offer several advantages. They provide higher photodetection efficiency (PDE) and, with a proper electronic readout, better time resolution. The last is not critical for standard VHE gamma-ray observations but can be relevant for other applications of IACTs such as intensity interferometry (\citep{intint_2020}). They do not experience any ageing when exposed to bright environments. This is particularly important because typically IACTs are either not operated under moonlight or must do hardware interventions that degrade the sensitivity of the instruments to operate the PMTs under relatively bright environments~\citep{MoonPerformance_2017}. The use of SiPMs in IACTs could significantly boost their duty cycle (up to a factor $\sim$2). SiPMs are much more compact and do not operate under high voltage, which a priori allows to reduce the cost, size and weight of a camera, which is critical for next-generation experiments in VHE gamma-ray astronomy. The future Cherenkov Telescope Array (CTA) will have more than 100 telescopes~\citep{Acharya_2013} and as it will be shown in the next sections, a large fraction of them will be equipped with SiPMs. Reducing the cost of a single pixel can be crucial for the realization of other proposed experiments like MACHETE~\citep{Machete_2016}, which requires a very large camera with many thousands of pixels.

SiPMs have some disadvantages which will be discussed in section~\ref{sec:SiPMsIACT}. Probably the main drawback (not only for VHE gamma-ray astronomy applications) is their limited physical size: they are rarely commercially available in sizes larger than 6$\times$6 mm$^{2}$. This constrains their employment in large cameras in which a reduction of a pixel size would dramatically increase the cost and complexity of the electronic readout. Different approaches to build large SiPM-based pixels for VHE gamma-ray astronomy have been developed and will be introduced in section~\ref{sec:LST}.

In this work we briefly review different SiPM developments in the context of VHE gamma-ray astronomy IACTs. For further details on these developments we refer the reader to their corresponding references. In section~\ref{sec:SiPMsIACT} we discuss the key-performance characteristics of SiPMs developed for Cherenkov astronomy. Section~\ref{sec:FACT} introduces FACT, the first IACT equipped with SiPMs. In section~\ref{sec:CTA} we introduce the different prototype telescopes of CTA that are using SiPMs and developments towards upgrading the cameras, based on PMTs, of current telescopes.

\section{SiPMs for Cherenkov astronomy}\label{sec:SiPMsIACT}

\begin{figure}[h]
    \centering
    \includegraphics[width=0.6\textwidth]{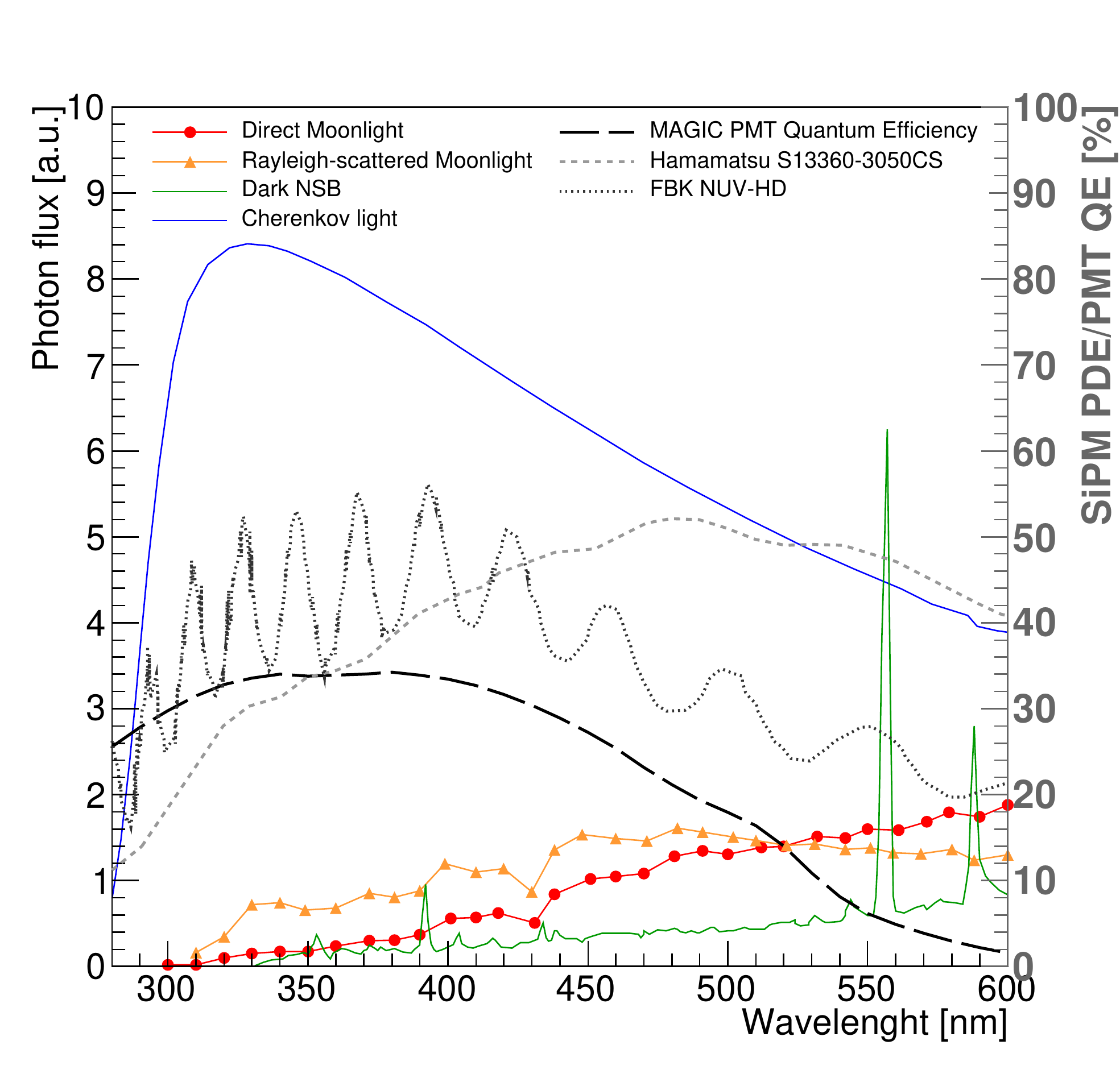}
    \caption{Modified version of Figure~2 in ~\citep{MoonPerformance_2017}  (Copyright 2017, reproduced with permission from Elsevier). The blue curve shows the typical Cherenkov light spectrum for a vertical shower initiated by a 1~TeV gamma ray, detected at 2200~m a.s.l.~\citep{1TeVshower}. In green, the emission spectrum of the NSB in the absence of moonlight measured in La Palma, Spain~\citep{Benn98}. In red, the shape of direct moonlight spectrum. In orange, the Rayleigh-scattered moonlight spectrum. The three curves were scaled by arbitrary normalization factors. The long-dashed black line shows the quantum efficiency of a PMT of the MAGIC telescopes. Gray dotted and short-dashed lines show the PDE of FBK and Hamamatsu SiPMs, respectively, taken from Figure~10 in~\citep{Otte_2017}.}
    \label{fig:pde}
\end{figure}

The Cherenkov light induced by VHE gamma rays in the atmosphere typically peaks at $\sim$350~nm (Figure~\ref{fig:pde}). Noise is dominated by the ambient light, which in the absence of moonlight and clouds is mainly produced by stars and dust, normally referred as \textit{dark} Night Sky Background (NSB). NSB peaks towards the red part of the visible band with typical rates of hundreds of MHz per pixel~\citep{Fink_2016}. Noise rates significantly increase when the Moon is above the horizon. The contribution from dark counts to total noise is then secondary. Actually, dark counts have proven to be useful during on-site calibration (cameras are kept closed during a dark-count-based calibration)~\citep{HAHN_2018}. In this section we briefly summarize the key aspects of SiPMs from the VHE astronomy point of view. For a detailed study on the performance of different SiPMs suitable for IACTs we refer the reader to~\citep{Otte_2017} and~\citep{ASANO_2018}.

An IACT photodetector should ideally have high sensitivity below 400~nm, but not so high at longer wavelengths where NSB and moonlight dominate. Traditionally SiPMs were actually not so sensitive in the near UV (NUV) band, and too sensitive in the green band. SiPMs with enhanced sensitivity in the blue and NUV bands were developed when these sensors became popular among the high-energy physics, astrophysics and medical physics communities. As shown in Figure~\ref{fig:pde} FBK NUV-HD~\citep{UVFBK_2017} SiPM exhibit a PDE curve that approximately follows the shape of typical Cherenkov pulses. New SiPMs can provide a PDE higher than 50\% at $\sim$350~nm, while keeping optical cross-talk close to $\sim5\%$, a regime in which its effect is sub-dominant with respect to the coincidence probability of two NSB events.

Camera temperature should be stable for a reliable operation of SiPMs. Tests performed with SiPM installed in one of the cameras of the MAGIC telescopes resulted on temperature variations below $1^\circ$C during the observation of a single source and below $1.5^\circ$C during a full night~\citep{HAHN_2018}. As introduced before, probably the main drawback of SiPMs is their limited physical size, especially problematic for building large cameras. Building SiPMs larger than 6$\times$6 mm$^{2}$ is normally not considered as a feasible solution mainly because capacitance significantly increases with size. Different solutions to build  SiPM-based large pixels for IACTs have been proposed and are discussed in section~\ref{sec:LST}.

\section{FACT, the first Cherenkove telescope using SiPMs}\label{sec:FACT}

The First G-APD\footnote{Geiger-avalanche photodiode} Cherenkov telescope (FACT) is the first IACT equipped with SiPMs and is operative in the Canary Island of La Palma, Spain, since 2011~\citep{FACT_2014}. The 53~cm diameter camera has 1440 pixels, each of them consisting of a $3\times3$~mm$^2$ SiPM coupled to a solid light concentrator that allows to reduce the dead area between pixels and acts as shielding for the light not coming from the reflector. The SiPMs are operated at a gain of $7.5\times10^5$, providing a peak PDE of $\sim$33\% between 450 and 550~nm and a crosstalk probability of~13\%. The dark count rate per pixel is of the order of a few MHz, well below the NSB rate.

\begin{figure}
    \centering
    \includegraphics[height=4cm]{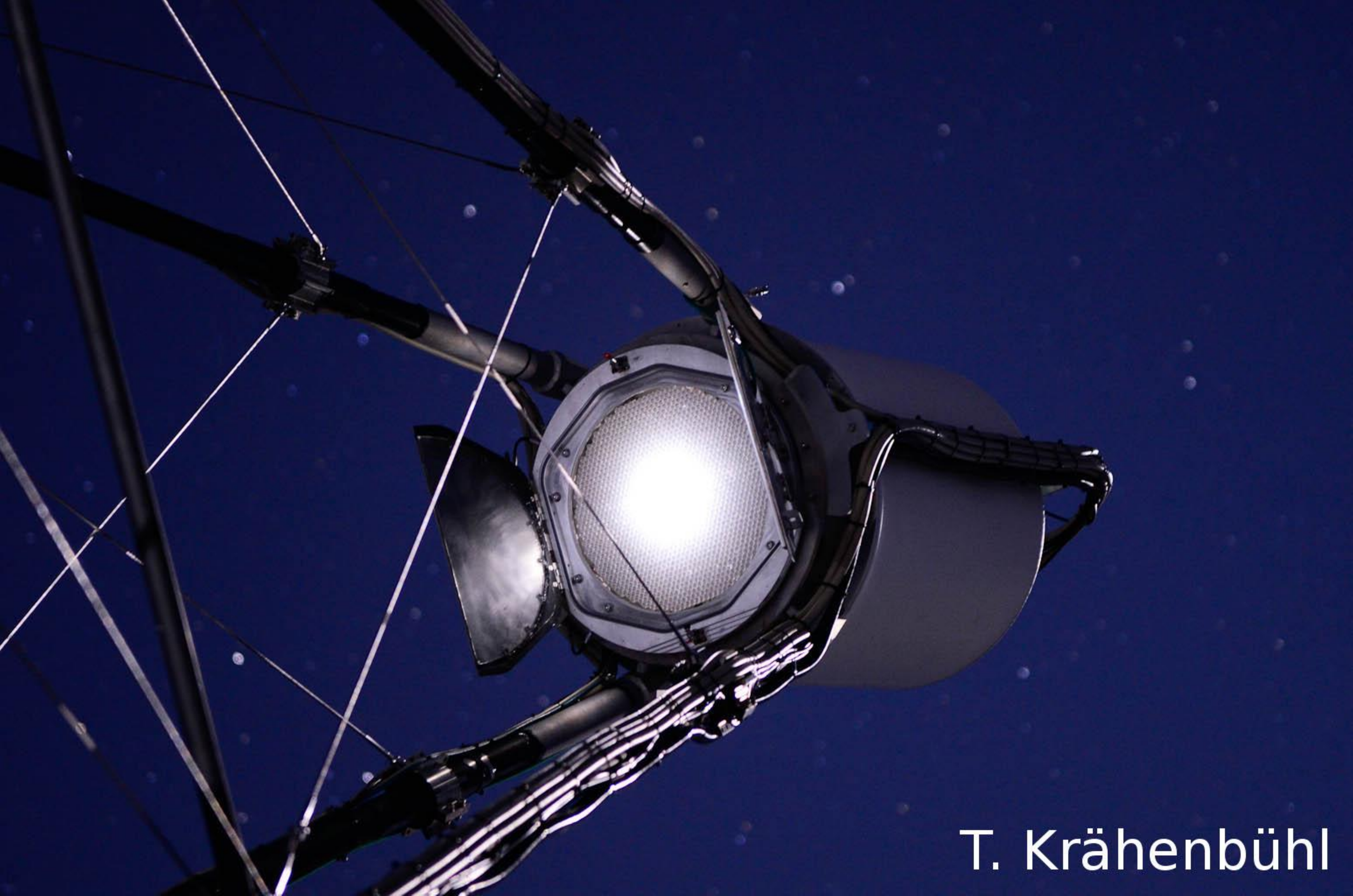}
    \includegraphics[height=4cm]{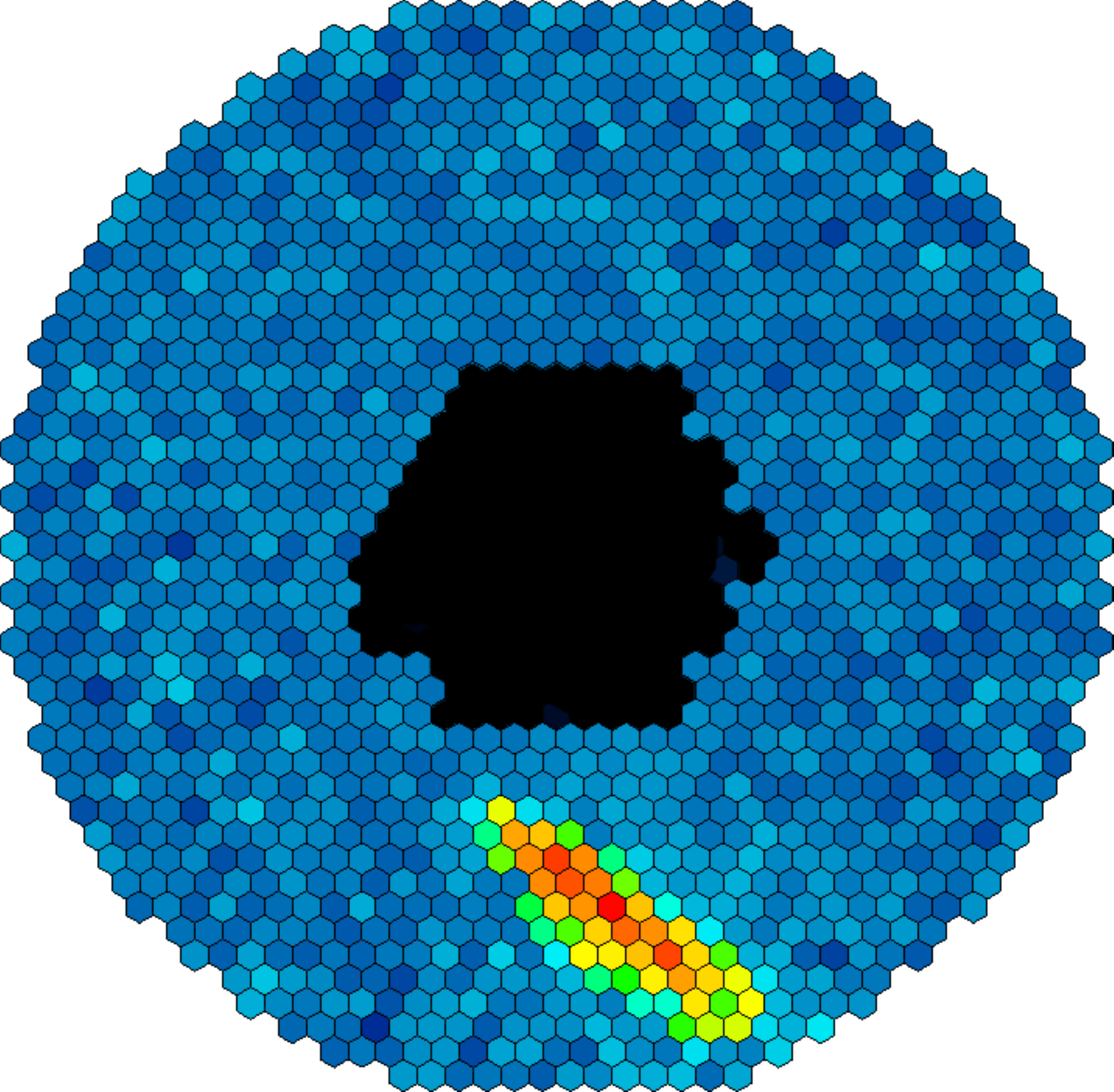}
    \caption{\textbf{Left:} an image of the FACT camera tracking the full Moon. \textbf{Right:} an event recorded while tracking the full Moon. The central pixels were disabled during data taking. Both images were obtained from~\citep{FactMoon}. Reproduced with permission from the author.}
    \label{fig:factmoon}
\end{figure}

One of the main goals of FACT was to prove the use of SiPMs in VHE gamma-ray astronomy, which was successfully accomplished. They were able to keep SiPM gain under control despite their temperature dependence, even with a photosensor technology that is now more than ten years old. Moreover, they were able to prove that SiPMs can be operated under strong moonlight~\citep{FactMoon} and even reconstruct shower images while pointing the telescope to the full Moon (see Figure~\ref{fig:factmoon}).

\section{SiPMs in CTA}\label{sec:CTA}

 The CTA observatory consists of two sites, one in the Northern and one in the Southern hemisphere. Three sizes of telescopes will be employed: small- (SST), mid- (MST) and large-size (LST) telescopes, targeting different energy ranges. The first telescope prototypes have already been built and are now under their commissioning phase. Two SSTs and one MST prototypes are equipped with SiPMs.

\subsection{SiPMs in SSTs}\label{sec:SST}

70 SSTs are planned to be built in the Southern hemisphere, targeting the highest energies (from a few TeV to $\sim$300~TeV). Three different designs were proposed and prototyped, the SST-1M~\citep{SiPM-SST1M} and ASTRI~\citep{SiPM-ASTRI} using SiPMs and the GCT~\cite{GCT} using multi-anode PMTs.

\begin{figure}
    \centering
    \includegraphics[height=5cm]{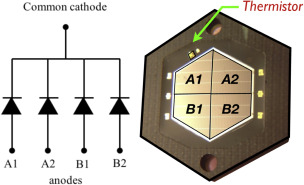}
    \includegraphics[height=5cm]{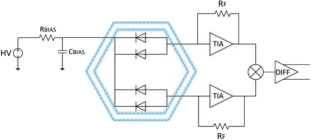}
    \caption{The hexagonal SiPM S10943-2832(X) that was develop and characterized in~\citep{Nagai_2019} (Copyright 2019, reproduced with permission from Elsevier) to be used in the SST-1M camera.}
    \label{fig:hexSiPM}
\end{figure}

The SST-1M camera is 0.9~m diameter and is composed of 1296 hexagonal SiPM pixels. The SiPM pixel, SiPM S10943-2832(X), is $\sim$10~mm flat-to-flat long and was developed by~\citep{Nagai_2019} in collaboration with Hamamatsu. The pixel is organized in quadrants (Figure~\ref{fig:hexSiPM}) in order to reduce the capacitance that such a large sensor would have if built following the prescription for smaller sensors. It has four independent anodes and a common cathode, allowing to readout the 4 channels independently while providing a single bias for the whole sensor. The four channels are summed in two steps to reduce the equivalent capacitance and pulse length.

\begin{figure}
    \centering
    \includegraphics[width=0.4\textwidth]{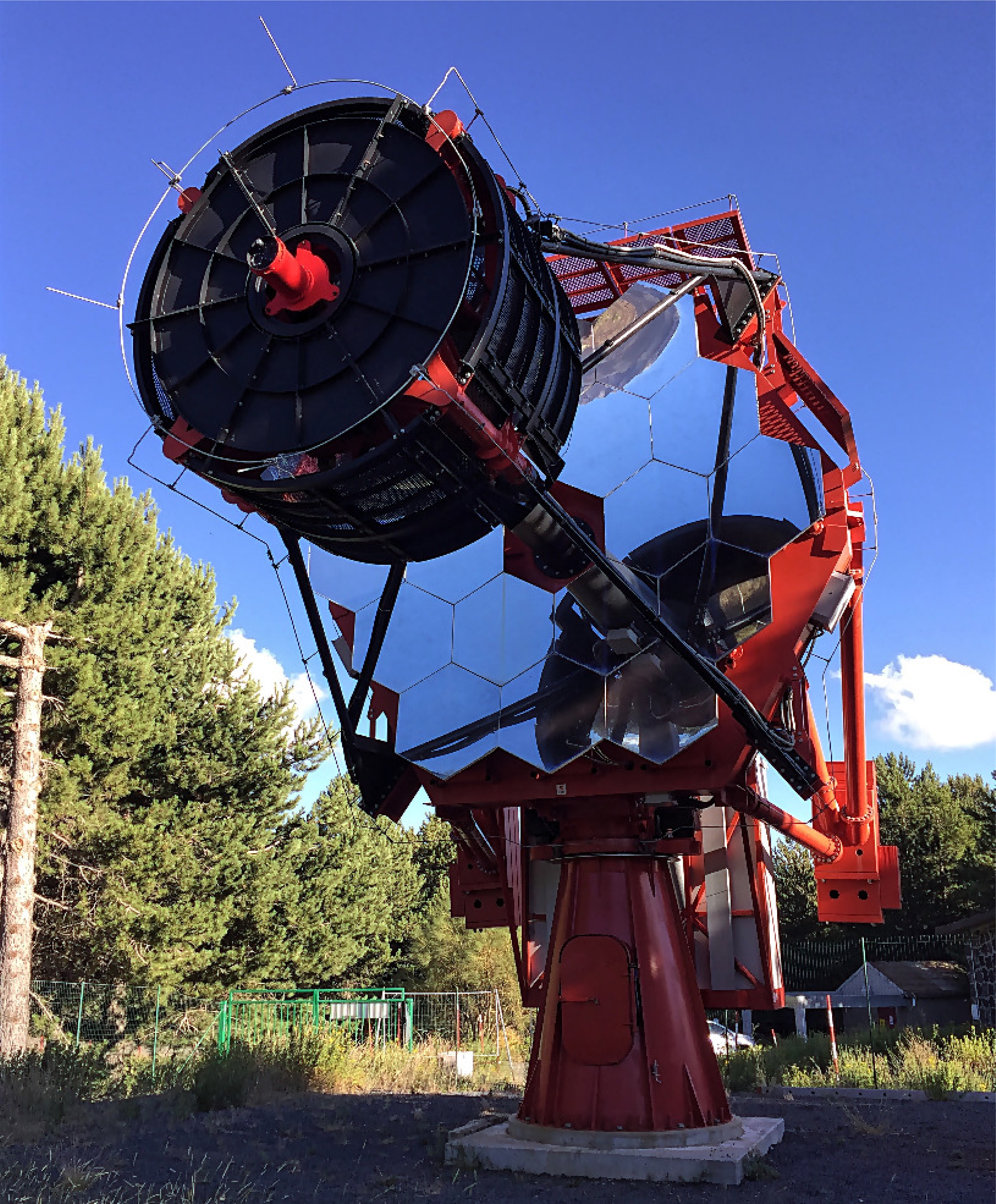}
    \caption{The ASTRI telescope and its dual mirror optics~\citep{Astri_2020}. Reproduced with permission, Copyright ESO.}
    \label{fig:astri}
\end{figure}

Traditional IACTs, including the LST or the SST-1M in CTA, are designed using single mirror optics. A better focusing, a larger field of view and a much smaller plate scale can be achieved by using the Schwarzschild-Couder dual-mirror optics~\cite{schwarzschildCouder}, as done by ASTRI and GCT (Figure~\ref{fig:astri}). The smaller plate scale results on a smaller camera that is easier to populate with SiPMs. In ASTRI the  SiPM  sensors  are  organized in 37 modules of 64 pixels each, with a pixel being a $6\times6$~mm$^2$ SiPM. With 21 of the 37 modules populated, ASTRI was the first IACT to detect the Crab Nebula using dual-mirror optics~\citep{Astri_2020}. 

 Light concentrators are normally used in single-mirror IACTs to reduce dead space between pixels and to reject NSB. The last is achieved by designing the concentrators to accept only light that is being focused by the mirror, at incident angles to $\sim30^\circ$. This solution is not suitable for Schwarzschild-Couder telescopes, where photons arrive to the focal plane with incident angles up to 60$^\circ$~\citep{OTTE_2015}. In these telescopes, dead space is reduced by tightly packing the SiPM pixels. To improve NSB rejection ASTRI camera is equipped with an infrared filter that rejects most of the light with wavelengths above 550~nm and at the same serves as protection for the sensors~\citep{Catalano_2018}.

\subsection{The pSCT, an MST equipped with SiPMs}

The MSTs are optimized for observations in the energy range from $\sim150$~GeV to $\sim$5~TeV. The CTA MST has a 12~m diameter mirror that focus the light into a PMT camera. A proposed alternative to MSTs is the prototype Schwarzschild-Couder Telescope (pSCT), which has been constructed to prove dual-mirror optics design on the MST scale. The smaller camera that dual-mirror optics employs allowed to build a camera equipped with SiPMs also in an MST. The pSCT is currently located at the Fred Lawrence Whipple Observatory in Arizona, has been inaugurated on January 2019 and its commissioning is ongoing.  Currently the pSCT has 1600 $6\times6$~mm$^2$ SiPM pixels and a 2.68$^\circ$ field of view. Once upgraded, the camera will have 11328 pixels with an 8$^\circ$ field of view.

\begin{figure}
    \centering
    \includegraphics[width=0.6\textwidth]{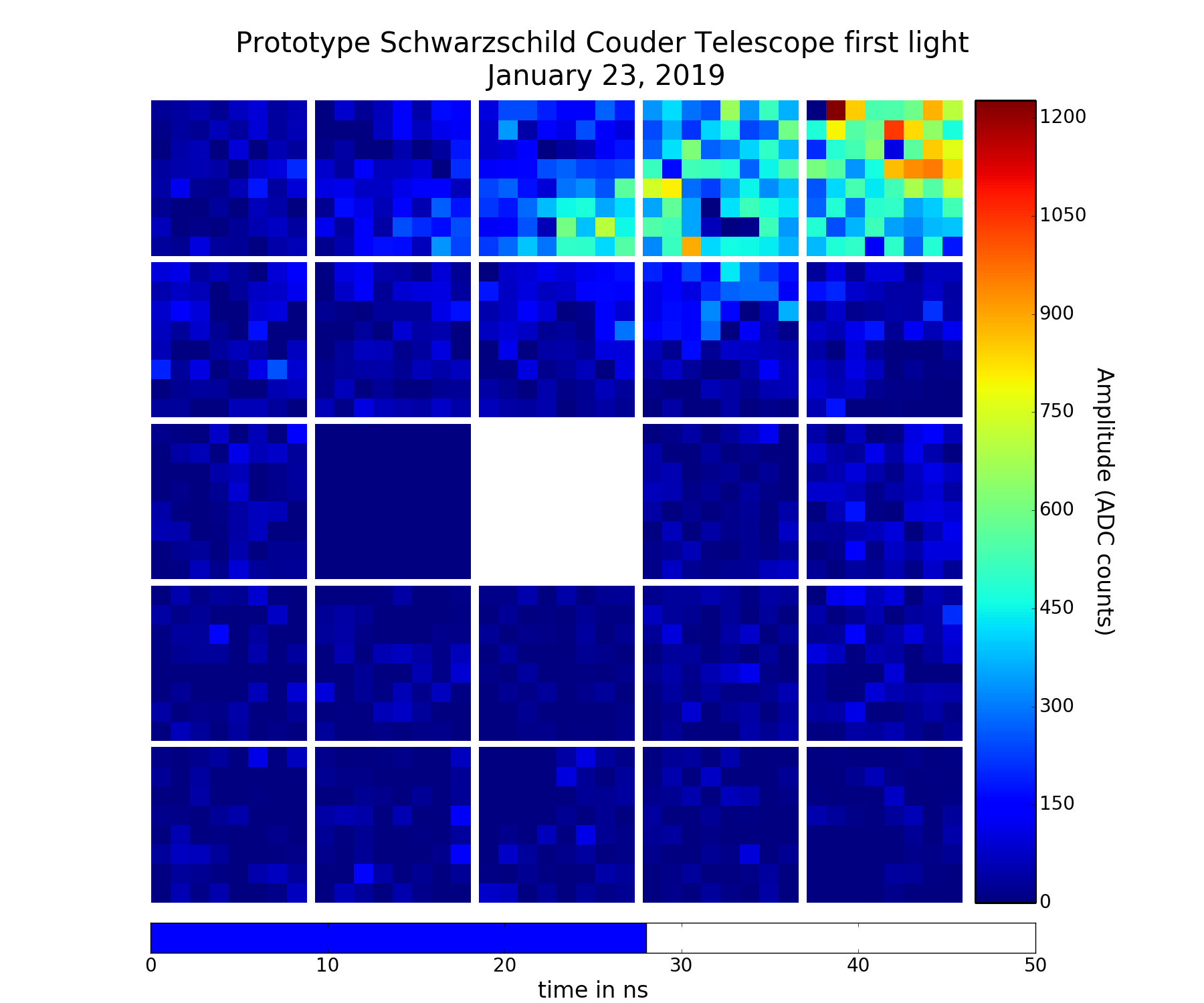}
    \caption{Single event recorded with the pSCT shown in~\citep{leslie}. The data used to produce this image did not undergo pedestal subtraction or data processing.}
    \label{fig:pSCT}
\end{figure}

Two types of sensors are populating the camera: Hamamatsu S12642-0404PA-50(X) and the newer 3rd generation NUV-HD of FBK. The Hamamatsu S12642 are $3\times3$~mm$^2$ SiPMs. In the pSCT 4 of them are connected in parallel to form a $6\times6$~mm$^2$ pixel. Compared to other Hamamatsu devices, they are operated at higher overvoltages, which reduces the dependence of gain an PDE on temperature. Since SiPM technology is rapidly evolving, Hamamatsu S126242 are relatively old devices with high crosstalk probability and a PDE that is not the most suitable for VHE gamma-ray astronomy (see~\citep{OTTE_2015} for the characterization of these sensors and a description of the SiPM temperature stabilization system of the pSCT camera). The FBK NUV-HD are $6\times6$~mm$^2$ SiPMs that achieve $\sim$60\% PDE at $\sim$350~nm~\citep{leonardoSPIE_2019}. A shower image recorded a few days after the pSCT inauguration is shown in Figure~\ref{fig:pSCT}. More details on the current status of the pSCT camera can be found in~\cite{leslie}.

\subsection{SiPMs for larger telescopes}\label{sec:LST}

The Large Size Telescopes (LSTs) feature the largest reflectors (23~m diameter) among CTA telescopes and target the lowest energies (with an energy threshold at $\sim$20~GeV)~\citep{LST_2019}. The first LST prototype was installed in 2018 and it is now going through the commissioning phase. Its camera dimensions are $2.9\times2.8\times1.15$~m$^3$ and holds 1855 PMT pixels, with a PMT diameter of 1.5". As mentioned in section~\ref{sec:intro}, pixel size is one of the main limitations to use SiPMs in these type of telescopes. However, different solutions to build large SiPM pixels for an eventual upgrade of the LST camera have been proposed.

\begin{figure}
    \centering
    \includegraphics[height=5cm]{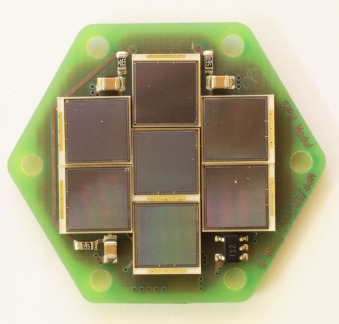}
    \includegraphics[height=5cm]{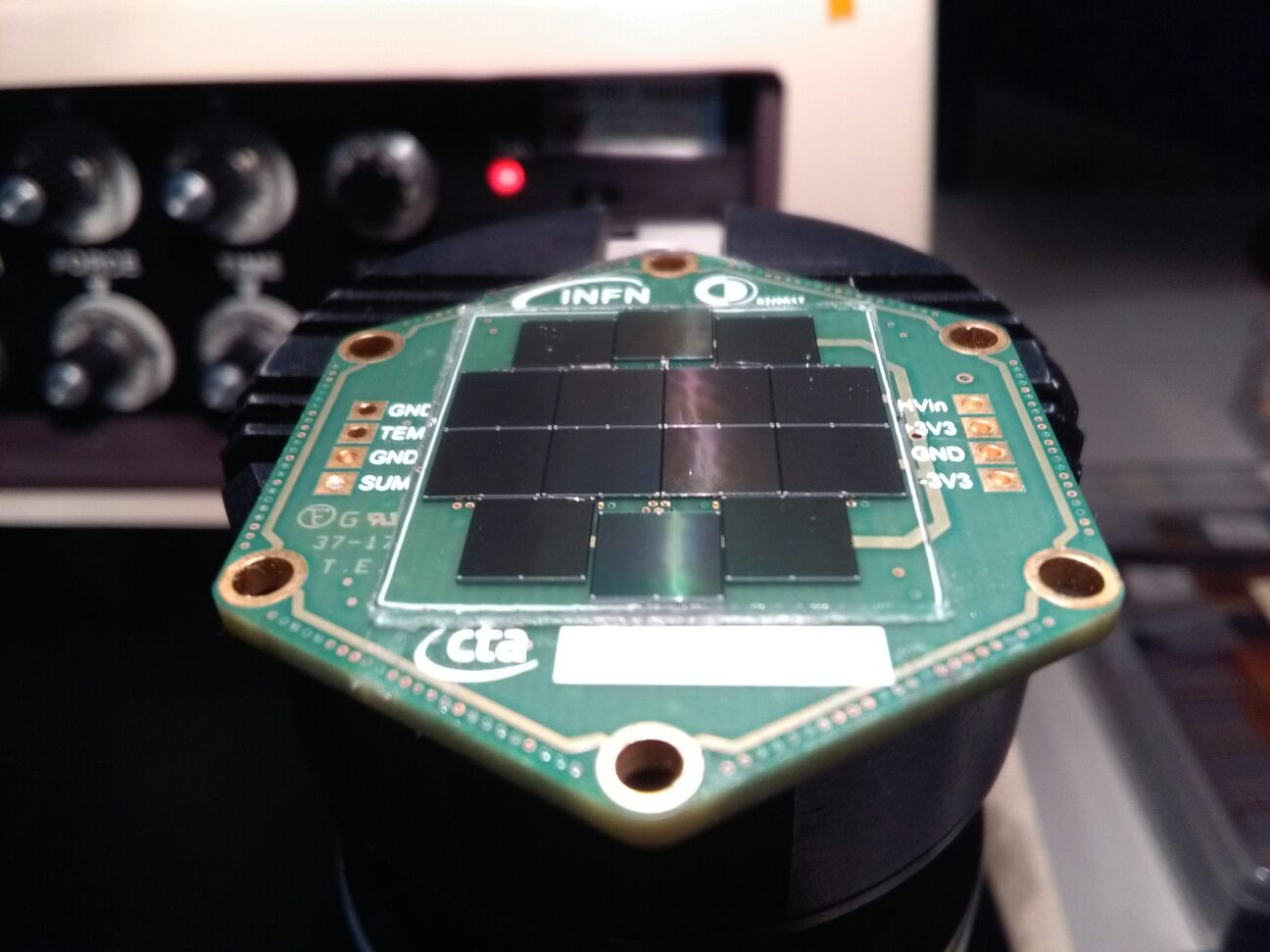}
    \caption{\textbf{Left:} First large SiPM pixel prototype installed in an IACT. It sums the output currents of 7 $6\times6$mm$^2$ Excelitas SiPMs and was installed in one of the edges of the MAGIC telescopes~\citep{HAHN_2018} (Copyright 2016, reproduced with permission from Elsevier). \textbf{Right:} Large SiPM pixel consisting on the sum of 14 SiPMs of 6$\times$6~mm$^2$ built by~\citep{Mallamaci_2019} (Copyright 2019, reproduced with permission from Elsevier).}
    \label{fig:SiPMLST}
\end{figure}

One promising approach is to build pixels made of several SiPMs ($\sim$10) tiled together (Figure~\ref{fig:SiPMLST}), where the output currents of the SiPMs are summed with operational amplifiers to output only one signal (reducing the number of readout channels needed by a factor $\sim$10)~\citep{Ambrosi_2016, HAHN_2018, Mallamaci_2019}. This way the dramatic increase in the capacitance is significantly diminished with respect to what would be achieved if connecting the SiPMs in parallel. However, capacitance still increases with the number of SiPMs that are being summed, meaning that this solution cannot be applied over a too large number of SiPMs if wanting to keep good single photoelectron and time resolution. In addition, the noise of all the SiPMs are being summed.

The MUSIC~\citep{Gomez_2016}, a multi-purpose application-specific integrated circuit (ASIC), was designed inspired by these developments. The possibility of having the pre-amplification and summation circuits inside an ASIC offers several advantages, like that it is easier to reproduce in a large scale and much more compact. One of the main functionalities of MUSIC is the possibility to perform the sum of up to 8 SiPMs. Different pole-zero configurations can be programmed to optimize the pulse shape for the application. Individual bias voltage offsets can be set to each SiPM and they can be even switched off, which can be particularly useful during calibration.

\begin{figure}
    \centering
    \includegraphics[width=\textwidth]{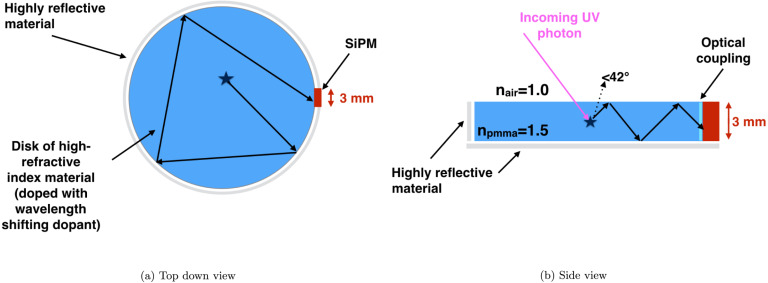}
    \caption{Conceptual design of the Light-Trap~\citep{Guberman_2019} (Copyright 2019, reproduced with permission from Elsevier).}
    \label{fig:LT}
\end{figure}

A completely different approach was reported by~\citep{Guberman_2019}. In the so-called Light-Trap pixel a single SiPM is coupled to a PMMA disk doped with a wavelength shifter (WLS). In this pixel the near UV photons of the Chrenkov flashes are absorbed by the WLS and re-emitted isotropically at longer wavelength where the SiPM PDE is higher. The wavelength-shifted photons are trapped inside the disk by total internal reflection (see Figure~\ref{fig:LT}). The sensitive area of the pixel is equal to the disk area, which can be tens of times larger than the SiPM area. The advantages of this solution is that it has the potential to be low-cost (down to che cost of a single SiPM), that capacitance does not increase with size (preserving good single photoelectron resolution), that there is no theoretical limit to the dimensions of the disk and that the pixel geometry can be easily modified. The main drawbacks are a degradation in the detection efficiency (a significant fraction of the wavelength-shifted photons escape the disk without reaching the SiPM) and in the timing properties of the pixel.

\section{Summary}\label{sec:conclusions}
Recent developments in SiPMs are challenging the hegemony of PMTs in VHE gamma-ray astronomy. Following the pioneer FACT, different IACTs with different mirror size and geometry that employ SiPMs have been built and are now being commissioned. It will be interesting to see how the advantages (higher PDE, possibility of operation under moonlight) and drawbacks (crosstalk, sensitivity to NSB) of SiPMs impact the performance of these telescopes. The development of large SiPM pixels operative at room temperature with reasonable capability to resolve individual photons remains as one of the main challenges in SiPM research, not only for IACTs. The possibility to use of SiPMs in large cameras depends on these developments.




\bibliography{mybibfile}






\end{document}